\documentclass[12pt,preprint]{aastex}
\def\p/{\mbox{$^1$}}
\def\pp/{\mbox{$^2$}}
\def\ppp/{\mbox{$^3$}}
\def\pppp/{\mbox{$^4$}}
\def\m/{\mbox{$^{-1}$}}
\def\mm/{\mbox{$^{-2}$}}
\def\mmm/{\mbox{$^{-3}$}}
\def\mmmm/{\mbox{$^{-4}$}}
\def\Ms/{\mbox{M$_\odot$}}
\def\MVmax/{\mbox{$M_{V,{\rm max}}$}}
\def\ha/{\mbox{H$\alpha$}}
\def\oiiir/{\mbox{[\ion{O}{3}]~$\lambda$5007}}
\def\re/{\mbox{$r_{1/8}$}}
\def\rq/{\mbox{$r_{1/4}$}}
\def\rh/{\mbox{$r_{1/2}$}}
\def\rqqq/{\mbox{$r_{3/4}$}}
\newcommand{\sci}[2]{\mbox{$#1\cdot 10^{#2}$}}
\newcommand{\scipm}[3]{\mbox{$(#1\pm #2)\cdot 10^{#3}$}}
\slugcomment{Accepted for publication in the Astrophysical Journal}
\begin{document}

\title{Structural Properties of Massive Young Clusters}     
\shorttitle{Structural Properties of Massive Young Clusters}

\author{Jes\'us Ma\'{\i}z-Apell\'aniz}
\affil{Space Telescope Science Institute\altaffilmark{1}, 3700 San Martin 
Drive, Baltimore, MD 21218, U.S.A.}


\altaffiltext{1}{The Space Telescope Science Institute is operated by the
Association of Universities for Research in Astronomy, Inc. under NASA
contract No. NAS5-26555.}

\begin{abstract}

	We have retrieved multicolor WFPC2/HST data from the STScI archive for
27 nearby Massive ($\gtrsim\sci{3}{4}$ \Ms/) Young ($< 20$ Myr) star Clusters 
(MYCs). The data represents the most-complete-to-date sample of clearly resolved
MYCs. We have analyzed their structural properties and have found that they
can be classified as either Super Star Clusters (SSCs) or as Scaled OB 
Associations 
(SOBAs). SSCs have a compact core possibly surrounded by a halo while SOBAs 
have no core. A morphological sequence can be established from SSCs with weak 
halos to SSCs with strong halos to SOBAs and we propose that this is linked to 
the original mass distribution of the parent giant molecular clouds. Our 
results indicate that a significant fraction of the stars in MYCs dissipate on 
timescales of 10 Gyr or less due to the extended character of some of the
clusters. Also, SSCs with ages $< 7$ Myr have smaller cores on average than 
those with ages $> 7$ Myr, confirming predictions of numerical simulations with 
mass loss. 

\end{abstract}

\keywords{galaxies: star clusters --- galaxies: starburst --- 
techniques: high angular resolution} 

\section{INTRODUCTION}

	Traditionally, star clusters have been classified as globular or open
clusters. Globular clusters are old ($\sim 10$ Gyr), massive 
($\sci{3}{4}-\sci{3}{6}$ \Ms/), metal-poor, and 
spherically-symmetric members of a halo population while open clusters are 
young ($\lesssim 1$ Gyr), low-mass ($< \sci{5}{3}$ \Ms/), metal-rich, 
and asymmetric members of a galactic disk. This classification is based on the 
cluster population of the Milky Way, where the dichotomy between the two types 
of clusters was first observed, but the situation is not the same for all 
galaxies. A clear example of this is the LMC, where we have $13-15$ classical 
($\approx 13$ Gyr old) globular clusters (GCs) but also
several intermediate age ($1-3$ Gyr old) and many young clusters \citep{DaCo01}.
Some of the young and intermediate LMC clusters have masses in the range 
$10^4-10^5$ \Ms/, similar to or slightly smaller than those of Galactic GCs
and are usually called in the literature ``rich clusters''
(see, e.g., \citealt{ElsoFall85}). Most of those rich clusters are part of the 
galactic disk \citep{Freeetal83}, so it is obvious that the traditional Galactic
classification cannot be directly extended to the LMC. Furthermore, in recent 
years it has been established that the Milky Way also has its own 
LMC-like rich clusters \citep{Moffetal94,Figeetal99b,Knod00}, so the 
traditional classification cannot be strictly applied even to the Galaxy.

	Other clusters which do not fit into the traditional 
classification are Super Star Clusters (or SSCs), first described in the nearby 
galaxies NGC 1569 and NGC 1705 by \citet{ArpSand85} and \citet{Melnetal85}. 
SSCs are compact Massive ($\gtrsim \sci{3}{4}$ \Ms/) Young 
Clusters\footnote{Note that, as is usually the case when classifying 
astronomical objects, no uniform definition of how massive a cluster has to be 
to be included in the ``massive'' category appears in the literature. Our 
choice is explained below.} (or MYCs) which, at first, could not be 
resolved from the ground and were mistaken for foreground stars. HST imaging 
\citep{OConetal94} was needed to establish their nature and to determine that 
they indeed belonged to their apparent host galaxies. Another interesting 
object is R136, the core of 30 Doradus in the LMC, which was once suspected to 
be a supermassive $1\,000-3\,000$ \Ms/ star until 
\citet{WeigBaie85} used holographic speckle interferometry to resolve it. Now 
it is clear that it is just another example of a compact MYC.

	The main reason why the traditional (Galactic-biased) classification 
does not include rich clusters or MYCs is that the Milky Way is not currently
in an active intense-star-formation phase, which accounts for the scarcity 
of those objects in the Galaxy. Furthermore, extinction at low latitudes 
severely hampers the detection of distant Galactic clusters. At the opposite
side of the activity spectrum from the Milky Way we have starburst and 
interacting galaxies such as the ``Antennae'' (NGC 4038/4039), where we observe
many clusters with masses larger than 10\pppp/ \Ms/ and ages 
less than 1 Gyr \citep{ZhanFall99}. In between, we have dwarf starburst galaxies
like NGC 4214 or NGC 5253, where several massive clusters with ages of less than
100 Myr are visible in the central regions. 

	Another problem with cluster classification is the distinction between
bound (or real) clusters and associations, which are unbound groups that are 
slowly dispersed by the galactic tidal field. Both types of objects are formed
from molecular clouds, with some of them being born as weakly bound clusters 
and later becoming associations due to mass loss and tides, thus complicating
the distinction\footnote{We are referring here to processes that
take place in Myr to Gyr time scales. In the long term all clusters
are expected to dissolve, maybe leaving a central black hole, but for 
massive clusters the time scales involved can be much longer than $10^{10}$
years and they can therefore be considered as bound in a first-order
approximation.}. It is only with the use of detailed kinematic
data that is it possible to differentiate between bound clusters and
associations, but such information is 
usually lacking for young extragalactic objects. Therefore, in this paper we
will use the term cluster in its broad sense of a stellar group formed from a
single cloud, and we will not assume that it is a bound object.

	What is the connection between massive young clusters and globular
clusters? Do all MYCs evolve to become GCs or do only a fraction survive after
10 Gyr? The keys to answer that question are the mass and the structure of the
cluster: only massive ($\gtrsim \sci{3}{4}$ \Ms/) clusters of the right size
(half-light radius $\approx 1-10$ pc) have a good chance of survival in a 
Hubble time scale \citep{FallRees77}. Therefore, we need to determine the 
distribution of masses 
and radii of MYCs in order to establish how many of them will become the GCs of
the future. As it happens most of the times in astronomy, it is easier to 
measure the total light output of an object than its mass, so most cluster mass
estimates are based on an assumed IMF or $M/L$ ratio. Thus, only a few SSC
masses have been directly measured \citep{HoFili96,HoFili01,GallSmit99,Lars01}, 
with the results (in the approximate range $10^5-10^6$ \Ms/) being consistent 
with the minimum required mass for survival. 

	In this paper we will concentrate on the second key by carrying an 
analysis of the radial structure of MYCs in order to answer the questions: 
Do all clusters have similar sizes? Do they all have similar core/halo mass 
ratios? How many clusters are likely to be bound? 
Previous studies have run into a problem: most galaxies with large 
numbers of MYCs are too far away to easily resolve the clusters, even with HST 
(e.g., the Antennae), so they had to deal with a large fraction of 
only-partially resolved objects. In other cases (e.g., M82), distance is not a 
problem but extinction is, since large-scale starbursts tend to be shrouded in 
dust. There is actually not a single example of a galaxy with a large sample of 
MYCs which is not affected by heavy extinction and which is located at a 
distance where present instrumentation can easily resolve the clusters. 
Therefore, if we want to build a well-resolved sample large enough to be useful 
to study the structural properties of MYCs we will need to compile our list 
from objects in several nearby galaxies.

	In section 2 we present our data describing the sample, showing our
measured values, and commenting on individual clusters. In section 3 we discuss
our results and in section 4 we present a summary.

\section{DATA}

\subsection{The sample}

	We have searched the HST archive for WFPC2 images of resolved MYCs in
nearby galaxies and selected the objects which matched the following criteria:

\begin{enumerate}
  \item	In order to be able to measure the structural properties of the objects
	in our sample without introducing any significant bias, we selected 
	only clusters within a radius of 5 Mpc (but see below for I Zw 18).
	This criterion eliminates, for example, the SSCs in He 2-10 
	\citep{Johnetal00} and the ``Antennae'' clusters \citep{Whitetal99b}.
  \item We selected MYCs with \MVmax/ (the age-corrected $M_V$, see below) 
	$< -11$, which eliminates some clusters like NGC 2363-B 
	\citep{Drisetal00}.  This criterion minimizes the confusion regarding 
	the extension of the cluster (dimmer objects are harder to distinguish 
	from the background and nearby clusters) and selects only the most 
	massive clusters. Indeed, a \sci{3}{4} \Ms/ cluster is expected from 
	evolutionary synthesis models \citep{Cervetal01,Leitetal99} to have an 
	\MVmax/ between $-11$ and $-12$ (the exact value depends on the 
	metallicity and the IMF at low masses), a prediction confirmed by the 
	measured masses (\citealt{HoFili01}). Thus, by using this criterion we 
	are sampling the likely predecessors of the GCs of the future.
  \item Only clusters with ages less than 20 Myr were included. This criterion
	eliminates clusters like NGC 4214-III \citep{Maizetal02a} and NGC
	5253-II and -III \citep{Calzetal97}. Older clusters tend to have more
	imprecise ages and values of \MVmax/.
  \item Dust-enshrouded clusters were also discarded due to the strong 
	geometrical distorsions induced by differential extinction around them.
	In this category there are highly-obscured objects like some of the 
	ones in M82 \citep{GallSmit99} and clusters with compact or 
	``filled'' nebular emission \citep{MaizWalb01} like NGC 4214-II-A and 
	II-B \citep{MacKetal00}, NGC 5253-V \citep{Calzetal97}, and NGC 2363-A 
	\citep{Drisetal00}. Those objects have moderate-to high extinctions 
	as well as strong nebular line emission and continuum with a compact
	spatial distribution which resembles that of the stellar continuum, thus
	severely hampering their distinction. The objects selected for
	our sample have low extinctions ($E(B-V) < 0.5$) with the only exception
	of the NGC 1569 clusters (where most of the extinction is of Galactic
	origin and rather uniform nature).
\end{enumerate}

	Twenty-seven clusters in eight galaxies were included in our sample. 
They are listed in Table~\ref{data}, along with the HST proposal IDs for the 
data. The clusters selected are very bright and in most cases they are among 
the most conspicuous optical structures in their host galaxies. We believe our 
data to be an over 50\% complete sample of low-extinction massive 
($\MVmax/ < -11$) young (age $< 20$ Myr) clusters within 5 Mpc. The largest 
degree of uncertainty is caused by the poorly known distances: some 
galaxies like NGC 1705 may actually be farther away than 5 Mpc while other 
ones which harbor good candidates (e.g. NGC 6946) may actually be closer.

	Given that most galaxies were observed under diverse proposals, the 
number and selection of filters available is very different for each of the 
clusters. In order to make the data more uniform, two optical bands were 
selected in each case: $U$ (F336W or F380W, but see below for the NGC 2403 case)
and $V$ (F555W or F547M, see below for the NGC 1705 case). The $U$ data will be 
our main source to study the structure of the clusters. We chose that band 
because it is the best optical tracer for the young stellar continuum, is little
affected by nebular contamination, has archival data available for all but one 
of our galaxies, and has the narrowest PSF. The differences between the two 
filters are small enough that no significant variations in the spatial structure
should appear. The $U$ images are displayed in 
Figs.~\ref{weak},~\ref{strong},~and~\ref{soba}.
The $V$ data will be used to obtain absolute magnitudes. In this 
sense it is preferred to the $U$ data because of its lower sensitivity to 
extinction.  F555W is very similar to the Johnson $V$ filter (for the age range 
of interest, the $({\rm F555W}-V)$ colors are $\approx 0.02$, 
\citealt{Holtetal95b}). F547M is narrower than F555W but is centered at a very 
similar wavelength, so the measured colors should be almost identical in most
cases. It is actually preferred to F555W due to its much lower sensitivity to 
nebular contamination. When strong nebular contamination was suspected in an
F555W exposure, F502N (\oiiir/) and/or F656N (\ha/) WFPC2 images were used to 
eliminate this effect.

	We compiled from the literature the available information regarding 
galaxy distances and cluster extinctions and ages (see Table~\ref{input}). 
Five different methods are
used in the available references to establish ages: color-magnitude 
diagrams, UV spectroscopy, optical spectral features (WR bands, Ca triplet), 
nebular equivalent widths, and integrated colors. Additionally, we used the data
presented by \citet{MaizWalb01} to detect the existence of \ha/ shells around 
the clusters and measure their sizes, thus placing an additional constraint on 
their ages: $\sim 2$ Myr old clusters have small ($\sim 10$ pc) shells around 
them, $\sim 4$ Myr old clusters have larger ($\sim 100$ pc) and usually broken 
shells, and for older clusters only diffuse \ha/ emission can be detected.
Finally, we also measured the integrated colors in our data and used the
\citet{Cervetal01} models\footnote{Available from 
{\tt http://www.laeff.esa.es/\~{}mcs/model} .} to produce a self-consistent age
and extinction \citep{MaizCerv01}. The final values for the ages and 
extinctions have quite different uncertainties depending on the number and 
quality of the sources. The distances to the Local Group clusters are 
quite precise ($5-10\%$ uncertainties) while the rest are not so well known 
($\sim 25\%$ uncertainties).

\subsection{Results}

	For each of the clusters, we measured the integrated $m_V$. 
We used the \citet{Cervetal01} models to correct for age differences by
calculating the dimming between an age of 4 Myr (the approximate age at which a
cluster attains its maximum optical brightness) and the current age. We then
introduced the values of the distance and extinction to calculate \MVmax/, the 
$M_V$ of the cluster at an age of 4 Myr. The values are listed in
Table~\ref{results}. No attempt has been made to estimate individual errors for 
\MVmax/, but for clusters outside the Local Group the typical uncertainties are
of the order of 1 magnitude, with the dominant source being the uncertainty in
the distance and, to a lesser degree, the age. The measurement uncertainties in 
$m_V$ itself are only relevant for Local Group objects, where the total errors
can be estimated as being less than 0.5 magnitudes.

\begin{figure}
\plotone{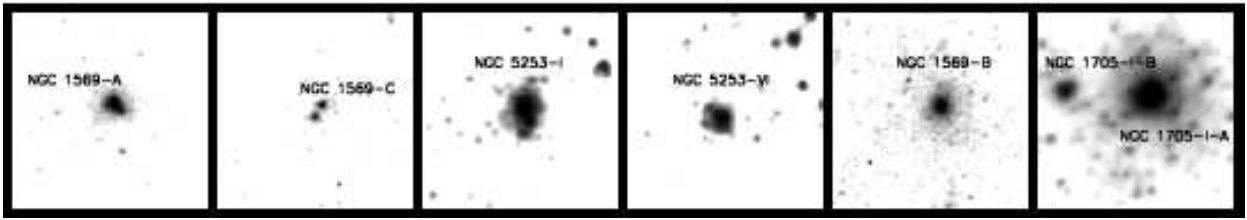}
\caption{F380W (NGC 1705 field) or F336W (rest) WFPC2/HST images of the compact
clusters in the sample with weak halos. The images have been 
resampled in order to use the same linear scale in all cases, with the field 
sizes being 50~pc~$\times$~50~pc. The orientation in each case is that of the
original archival image.}
\label{weak}
\end{figure}

\begin{figure}
\plotone{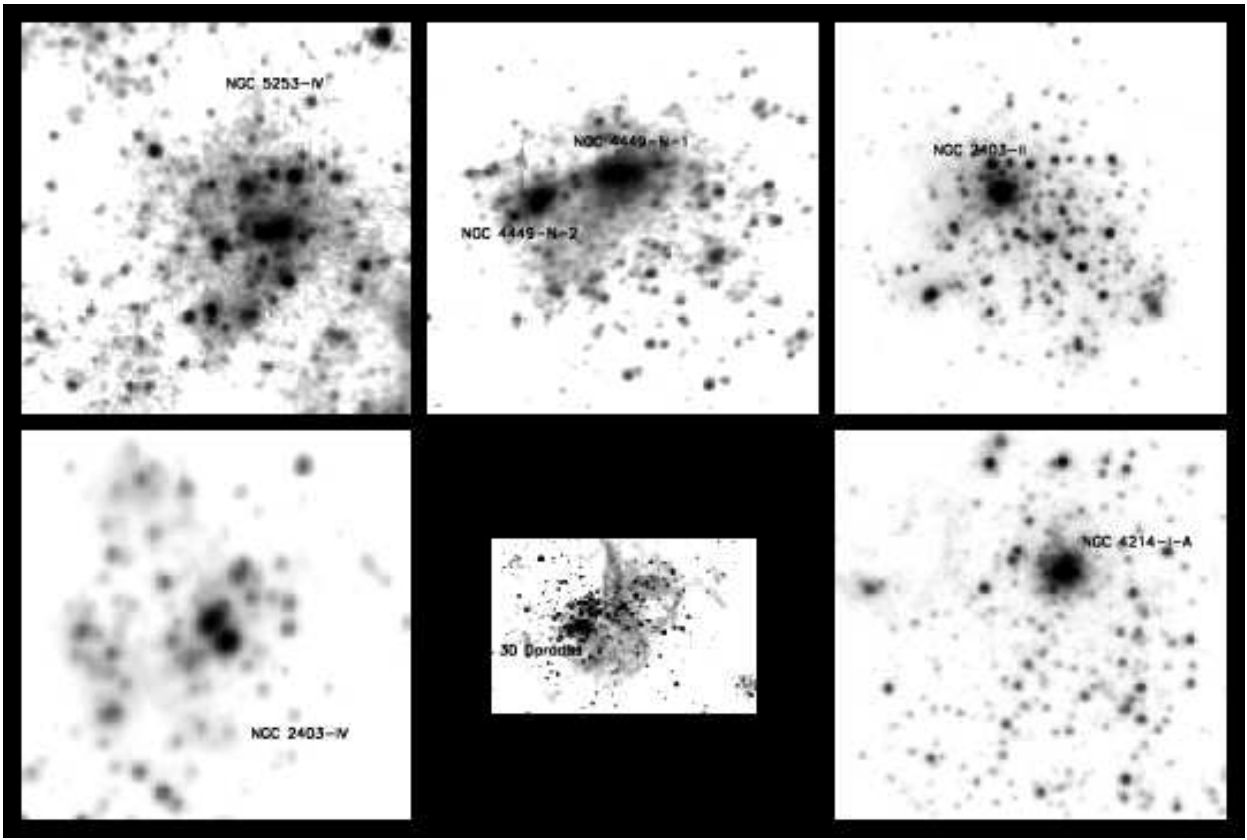}
\caption{Same as Fig.~\ref{weak} for the compact clusters in the
sample with strong halos. The NGC 2403 fields were obtained with the F439W
filter. The linear scale is the same and the field sizes are 
100~pc~$\times$~100~pc, except for the 30 Doradus field, which is 
68~pc~$\times$~45~pc.}
\label{strong}
\end{figure}

\begin{figure}
\plotone{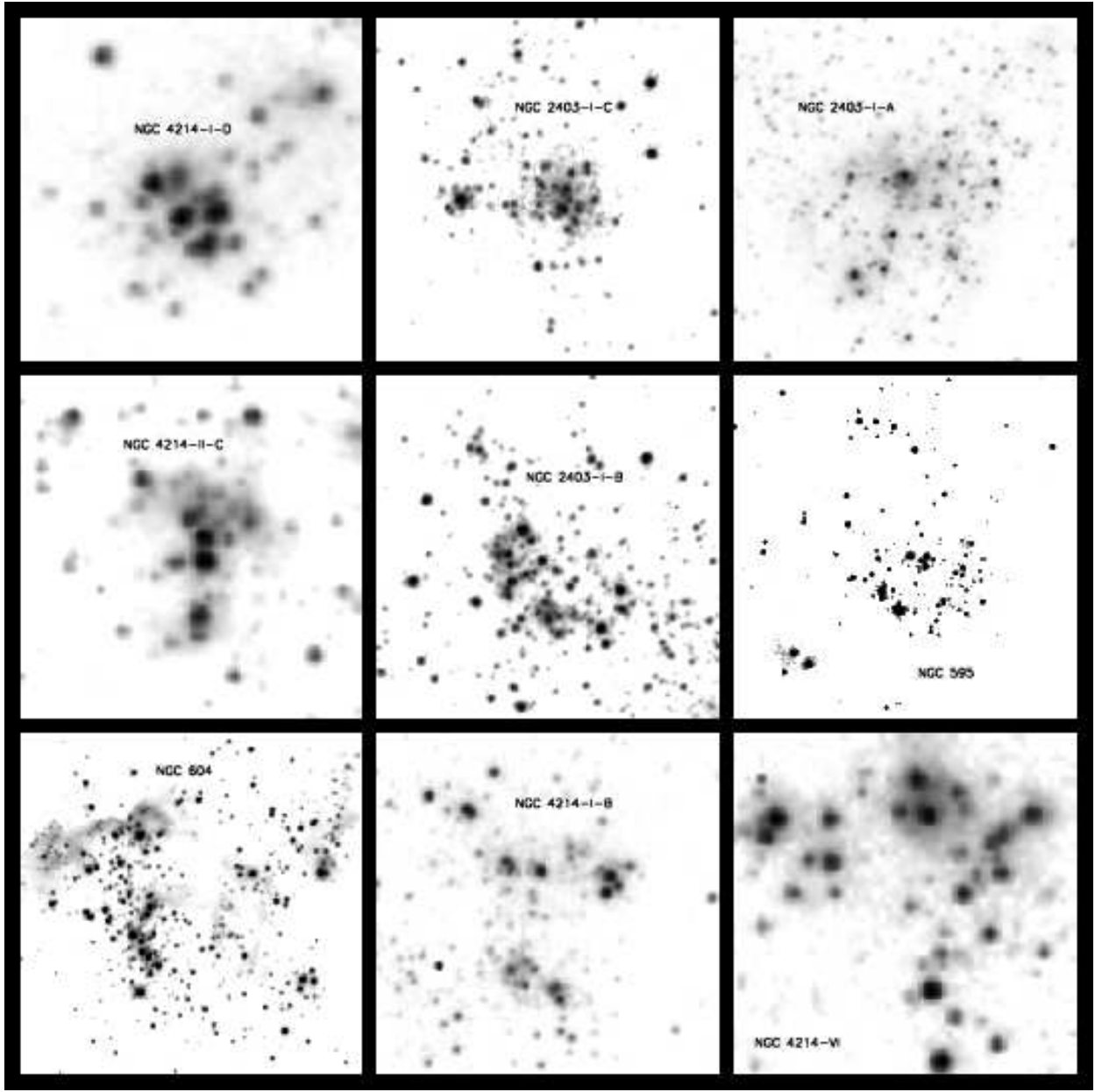}
\caption{(a) Same as Fig.~\ref{strong} for the SOBAs the sample. The field 
sizes are 100~pc~$\times$~100~pc.}
\label{soba}
\end{figure}

	Analyzing the $U$ band images we discovered that some clusters have a
distinct compact core (Figs.~\ref{weak}~and~\ref{strong}) while in others no
core is readily apparent (Fig.~\ref{soba}). Compact cores are easily 
distinguished because they have integrated values of $\MVmax/ < -10$ within a 
radius of $\lesssim 3$ pc (see Table~\ref{results}), which is $\gtrsim 2$ 
magnitudes brighter than the most luminous surrounding stars. Most cores show 
an approximate circular symmetry but some are double or elongated. Massive 
clusters with no core have the appearance of an OB association in terms of shape
and size but they are much more massive than the known Galactic ones. We 
propose here the use of the term ``Scaled OB Association'' (or SOBA) to refer 
to them, following a previous suggestion by \citet{Hunt99}. SOBAs are quite 
asymmetric extended objects with no well-defined center\footnote{This lack of 
a well defined center makes the values of \MVmax/ within 3 pc listed in 
Table~\ref{results} for SOBAs somewhat arbitrary, with a different choice of
aperture probably producing results different by up to 0.5 magnitudes.
Note, however, that with the only exception of the I Zw 18 clusters we always
tried to center our apertures at the brightest point source, so other choices 
would only probably make \MVmax/ within 3 pc dimmer.} and are likely to be 
weakly bound, if bound at all.

	One difference is also readily apparent between the clusters with a 
compact core (or compact clusters, for short) in Fig.~\ref{weak} and those in 
Fig.~\ref{strong}: The first ones do not have a halo or only a weak one around 
the core while for the second ones the halo is as luminous or even more so than
the core itself. We will refer to the clusters in Fig.~\ref{weak} as compact
clusters with weak halos and to those in Fig.~\ref{strong} as compact clusters
with strong halos. Halos have a structure similar to that of SOBAs
(Fig.~\ref{compintegrint}) and their approximate center usually does not 
coincide with the cluster core.

	We would like to test whether this morphological classification
corresponds to real structural differences or not. In particular, we would like
to know whether the core-halo structure of the objects in Fig.~\ref{strong} is
caused by the existence of two distinct structural components or whether the
core is just the central region of an extended one-component structure. A
way to test this would be to try to fit a King profile \citep{King62} in
each case. Unfortunately, this is not possible due to the diverse spatial
resolution of the data: even though all cores are resolved in the sense of being
at least significantly broader than nearby stars, they are not always resolved
to the point of being able to unambiguously measure the profile parameters. 
However, we can settle for a poor-man's version of this procedure by measuring
the $U$-band \rq/, \rh/, and \rqqq/, the one-quarter, one-half, and
three-quarters light-radii\footnote{Note that in a preliminary version of this
work \citep{MaizWalb01}, \rh/ was defined as applying only to the cluster core 
in the case of a compact cluster. Here, \rq/, \rh/, and \rqqq/ refer to the 
whole cluster always.}, and constructing the ratio 
$\alpha = \rh/^2/(\rq/\rqqq/)$.  Single-component King profiles with
reasonable values of $r_t/r_c$ (the ratio of the tidal to the core radii) 
are expected to have values of $\alpha$ very
close to 1.0, as shown in Fig.~\ref{classifplot}. On the other hand, clusters
with two components with $\lesssim 50\%$ of the light originating in the 
compact one and $\gtrsim 50\%$ in the extended one should have values of \rq/ 
(determined fundamentally by the compact component) smaller than expected for 
given ones of \rh/ and \rqqq/ (determined fundamentally by the extended 
component), which should lead to values of $\alpha$ larger than one. A problem
associated with these measurements (or with any other measurement of cluster 
brightness profiles) is the definition of the total light radius, which is 
needed for a correct subtraction of the background. This was done by selecting 
an initial guess from a visual inspection of the image and then modyfing it
until the area just outside the total light radius had an approximately flat
intensity and color radial profile. The adjacent area was then used to 
determine the background. When two clusters were close to each other, masks were
used to avoid mutual contamination.

	In Table~\ref{results} we show the measured values for \rq/, \rh/, 
and \rqqq/. The \rq/ values have been corrected for the finite value of the PSF
width but the corrections turned out to be unimportant in most cases.
The values of $\alpha$ as a function of \rh/ are plotted in
Fig.~\ref{classifplot}. There we can see that compact clusters with weak halos
(with one exception, NGC 1569-C, one of the clusters with a double core) and
SOBAs are reasonably well adjusted to the prediction of a single component King
model ($\alpha \approx 1$). However, compact clusters with strong halos all 
have values of $\alpha$ clearly greater than 1, indicating that they are indeed
made out of two different structural components. Furthermore, the fact that the 
three cluster classes occupy different areas of this $\rh/-\alpha$ diagram 
reinforces the reality of the differences suggested by the morphological 
classification. 

	We can then conclude that the MYCs in our sample are made out of two 
structural components: compact cores with half-light radii of $< 5$ pc and 
extended halos with half-light radii $> 15$ pc. In some cases one of the two 
components is absent or weak, and then we have a compact cluster with a weak 
halo or a SOBA. In other occasions both have a significant contribution and the 
result is a compact cluster with a strong halo. However, no single-component 
clusters with half-light radii of $\sim 10$ pc are detected in our sample, as
evidenced by the ``hole'' around $\rh/ \approx 10$ pc, $\alpha \approx 1$ in
Fig.~\ref{classifplot}. 

	Other studies \citep{Meuretal95,Whitetal99b} have not detected the
compact cluster-SOBA dichotomy. Besides the problem with spatial resolution 
at large distances and the mixture of high-mass and low-mass clusters, there is
another explanation for this lack of detection. As shown in Fig.~\ref{r14r12},
when \rh/ is used as a measurement of cluster size, compact clusters with
strong halos appear as intermediate size objects which fill the gap between the
other two classes, so the size histogram does not
show a strong bimodality. However, when \rq/ is used, the two peaks are quite
clear. This is explained by the fact that for compact clusters \rq/ is 
determined fundamentally by the core and is more or less independent of the
strength of the halo (though it may depend on whether the core itself is double 
or elongated).

\subsection{Object nomenclature and notes}

{\bf 30 Doradus, NGC 595, and NGC 604:} These clusters are the three brightest
low-extinction MYCs in the Local Group. 30 Doradus is in the LMC 
and NGC 595 and NGC 604 are in M33. The core of 30
Doradus, R136, fits easily in the PC field of view 
\citep{Huntetal95} but the halo is much more extended and we had to use the 5 
WFPC2-fields mosaic generated by \citet{Walbetal01} in order to cover it. R136 
is the archetype of compact clusters with a strong halo, since $\approx 90\%$ 
of its total integrated light originates there. The images of 30 Doradus and 
NGC 595 shown in Figs.~\ref{strong}~and~\ref{soba}, respectively, were
convolved with a gaussian kernel for display purposes.

{\bf I Zw 18-I and -II:} These low-metallicity clusters are included even though
I Zw 18 is at a distance of 10 Mpc because their extended character makes them 
easy to resolve. They are sometimes called I Zw 18-NW and -SE, respectively.

{\bf NGC 1569 clusters:} NGC 1569-A has a double core, with the two components 
separated by only 2 pc in the plane of the sky and with an $\approx$ 1.3 
magnitude difference between them \citep{DeMaetal97}. The detection of both WR 
emission features and the near-IR Ca\,{\sc ii} triplet in absorption in their 
unresolved spectrum led \citet{GonDetal97} to suggest an age difference of 
$\approx 6$ Myr between the two. However, the measured color difference between 
the two cores is small (with some likely mutual contamination), so it is not 
straightforward to test this hypothesis with the available data 
\citep{DeMaetal97} and here we will treat NGC 1569-A as a single cluster. 
NGC 1569-A is the 
archetype of compact clusters with a weak halo, since only $\lesssim 5\%$ of 
its total integrated light originates there. NGC 1569-C was classified as 
number 10 by \citet{Huntetal00}. As pointed out by \citet{Bucketal00}, two 
cores can be identified (separation $\approx 4$ pc). Note the existence of a 
significant internal extinction compared to NGC 1569-A and -B. The anomalous
location of NGC 1569-C in Fig.~\ref{r14r12} (it is the only compact cluster with
a weak halo with $\alpha$ significantly greater than 1) is caused by the double
nature of its core.

{\bf NGC 1705-I-A and -I-B:} NGC 1705-I-A is the bright SSC described by
\citet{Melnetal85}. NGC 1705-I-B is the second brightest cluster in the galaxy
and is located at a projected distance of 1\farcs 0 (24 pc) from NGC 1705-I-A,
which we consider large enough to treat them as individual clusters. In 
\citet{OConetal94} NGC 1705-I-B is referred to as cluster 35 and in
\citet{Meuretal95} as NGC 1705-2. In most of the WFPC2 images 
of this galaxy available in the HST archive at the present time, NGC 1705-I-A is
saturated, and in the ones in which it is not saturated, a tracking problem 
produced an elongated PSF, making them useless for our purposes of measuring 
the radial intensity profile. Fortunately, the saturation in the two F380W 
images with correct tracking is weak, affecting only the central $3\times 3$ 
pixels. Thus, we used the integrated photometry from the unsaturated F380W 
image with incorrect tracking to correct the flux in those central 9 pixels 
(only a 26\% increase in the total number of counts was required). The measured
value of \rq/ (corrected for saturation but not for the width of the PSF) is
$1.44\pm 0.10$ pixels (compared to an \rq/ value for a point source of 
$\approx 0.4$ pixels), with the exact value depending on how the extra flux is
allocated inside the central 9 pixels. We can then conclude that the uncertainty
introduced by the saturation correction in the value of \rq/ for NGC
1705-I-A is tolerable (i.e. it is much smaller than the one introduced by the
uncertainty in the distance). Another problem we had to face was that
in all of the F555W images 
available in the HST archive NGC 1705-I-A and -I-B are saturated, so $m_V$ had 
to be obtained from \citet{OConetal94} and \citet{HoFili96}. Their values are 
consistent with our measured $m_{\rm F380W}$ and $m_{\rm F439W}$ and the known 
colors of NGC 1705-I-A.

{\bf NGC 2403 clusters:} We follow the nomenclature of \citet{Drisetal99b}.
NGC 2403-I-A shows a weak core but we classified it as a SOBA because the value
of \MVmax/ within 3 pc is dimmer than $-10$. NGC 2403-IV is a compact cluster 
with a double core (separation $\approx 7$ pc), as shown in Fig.~\ref{strong}. 
NGC 2403 is the only galaxy with no $U$ observations available, so the filter 
with the most similar characteristics, F439W (WFPC2 $B$), was selected to 
analyze the structure of its clusters. 

{\bf NGC 4214 clusters:} We follow the nomenclature of \citet{MacKetal00}.
NGC 4214-II-A and -B are not included in the sample because of strong nebular 
contamination. NGC 4214-III and -IV are excluded because they are older than 
20 Myr. We adopt a distance of 4.1 Mpc but it should be noted that a preliminary
analysis of WFPC2 stellar photometry (final results will be published as 
\citealt{Maizetal02a}) indicates that the distance could be up to a factor of 
$\sim 2$ smaller. Such a change would not alter our conclusions since its 
effects would be only to exclude NGC 4214-II-C and NGC 4214-VI from our sample 
(due to the $\MVmax/ < -11$ requirement) and maybe to change the classification
of NGC 4214-I-C.

{\bf NGC 4449-N-1 and -N-2:} These two objects are located in the nuclear region
(hence the designation N) and they are the two brightest clusters in the 
galaxy. In \citet{Gelaetal01} they are called 1 and 31, respectively. Both
cluster cores are separated by $\approx 22$ pc (enough to consider them as 
individual clusters) and have elongated shapes. They appear to be in the process
of being torn apart by tidal forces.

{\bf NGC 5253 clusters:} We follow the nomenclature of \citet{Calzetal97} but 
using roman numerals instead of arabic ones for consistency with the rest of
our numbered clusters. NGC 5253-V is not included in our sample due to its 
heavy extinction. NGC 5253-II and -III are also excluded because they are older 
than 20 Myr. NGC 5253-IV has a double core (separation $\approx 4.5$ pc).

\section{DISCUSSION}

\subsection{Consequences of the morphological dichotomy}

	The different sizes of the cores and halos (including SOBAs in this last
category) of MYCs and their similar masses imply that the orbital time scales 
must be very different. Thus, a star in a typical core has an orbital period of 
$0.3-1.0$ Myr while one in a typical halo has an orbital period of $10-100$ Myr 
(if it is bound at all). Given that the clusters in our sample are all very 
young, we can easily conclude that the distribution of stars in the halo (when 
present) must closely follow their original distribution at birth. Cores may
have experienced some evolution (as we will see later) but it is also evident 
that their existence must be the result of the formation process. A consequence 
of this early evolutionary state is that strong halos are quite asymmetric
and are not centered around their respective cores, as would be expected in 
very young clusters whose stars have not completed a single orbit around the
center. We propose here that the existence of SOBAs and compact clusters can be 
traced to the existence of two broad types of cluster-forming 
Giant\footnote{Of course, ``giant'' should be understood in this context as
implying a very large mass, not a very large size.} Molecular Clouds (GMCs): 
``Super-GMCs'', characterized by very large masses and sizes of tens to a few 
hundred pc, and ``Compact-GMCs'', smaller and maybe less massive but with 
higher central densities. A compact cluster with a strong halo would be the 
result of a ``Compact-GMC-like'' core inside a ``Super-GMC''. The observed 
morphology of the nearby cloud OMC-1 \citep{WiseHo98} strongly reminds of this 
type of arrangement: A compact core is surrounded by a series of linear 
structures or filaments which extend for tens of core radii (see
Fig.~\ref{starchains}). Of course, the progenitors of MYCs must be much 
larger than OMC-1 but more massive cores are also observed in
other clouds \citep{Evan99} and the hierarchical structure of OMC-1 
suggests that the same type of structure could be present in
larger GMCs. In this respect it is interesting to notice that ``chains of
stars'' are readily visible in some of the SOBAs and massive halos of some of
the youngest objects in our sample, such as 30 Doradus, NGC 604, NGC 2403-II,
and NGC 4214-I-A, suggesting that they originated from molecular filaments.

	It has been suggested that the most massive stars form by the
coalescence of lower-mass stars (see \citealt{Stahetal00} for a recent review). 
If that was the case, one should find that the fraction of very massive stars 
depends on the density of the cluster, since stellar collisions should be 
very rare in SOBAs but rather common in compact cores \citep{PorZetal99}.
However, the analysis of a SOBA like NGC 604 reveals no obvious dearth of very 
massive stars. \citet{Huntetal96} analyzed the stellar photometry of the cluster
and measured values for the IMF and the ratio of WR/O stars similar to those
of R136. \citet{GonDPere00} studied the integrated UV spectrum of NGC 604 and
concluded that the best fit was provided by a 3 Myr old burst with a Salpeter 
IMF and $M_{\rm up} >$ 80 \Ms/, with the $M_{\rm up} \le$ 60 \Ms/ models clearly
excluded. Therefore, either coalescence plays only a minor role in the formation
of massive stars or the molecular filaments which appear to be the origin of
SOBAs and halos contain compact subcores which are dense enough to cause a
significant number of stellar collisions but not large enough to produce a 
compact cluster.

	Another important consequence of the morphological dichotomy between
compact clusters and SOBAs is the long-term survival of MYCs. Several
processes contribute to the destruction of a cluster, of which the most
important ones are two-body evaporation and tidal disruption by encounters with
the galactic disk or by dynamical friction with the galactic halo 
\citep{Fall01}. Two-body evaporation dominates for very compact clusters while
tidal processes control the fate of extended clusters, with both effects
becoming more important for low-mass clusters. Tidal effects strongly
depend on the galactic environment (a cluster close to a galactic nucleus is
much more easily destroyed than one in a circular orbit in the outer halo) 
and the evolution of a cluster is controlled by stochastic events so it is not 
possible to give a precise mass-size survival range for a given age. However, 
an approximate rule is shown in Fig.~2 of \citet{FallRees77} for the range of 
interest: A cluster with $M\gtrsim \sci{3}{4}$ \Ms/ and \rh/ between 2 pc and 
10 pc has a good chance of surviving after a Hubble time, while one outside 
this range is likely to be destroyed.  
Thus, SOBAs will likely disperse and compact clusters will likely lose
their halos due to tidal effects. The fate of an object like R136 will depend on
what fraction of its halo is able to retain. If it loses most of it, it could 
slip below the critical mass but if it retains the inner part it could survive
for a Hubble time. In any case, SOBAs will not be able to continue existing for
such a period of time and their components will mix with the rest of their
parent galaxies. Therefore, a large fraction of MYCs are expected to disperse
and at most 50\% will be able to become the GCs of the future.

	As it has been mentioned before, some of the cluster cores are double.
There are two possible explanations for this: (1) A simple manifestation of the
original structure of the molecular cloud, with two large mass concentrations
instead of one collapsing simultaneously. (2) An initial collapse of one of the 
two concentrations followed by the induced collapse of the second one. The
second process is called a {\em two-stage starburst} and is observed in the 
two most massive clusters in the LMC, 30 Doradus and N11
\citep{Parketal92,Walbetal99}. In a two-stage starburst there is an age
difference of $\approx 2-3$ Myr between the two stages and the final outcome can
be a double cluster (N11) or a central core surrounded by a younger generation
in its halo (30 Doradus). 
It would be interesting to obtain resolved spectroscopy of the 
clusters in the sample in order to decide whether double cores (or core-halo
structures) are coeval or whether there is an age difference between them.

	Finally we would like to discuss an aspect of the nomenclature which is
affected by the previous discussion. In the last years, the term {\em Super Star
Cluster} has become popular and it is applied to the progenitors of ``old''
globular clusters. In this paper we have avoided its use so far because, as we
have seen, it was not clear whether all MYCs are expected to become GCs and,
indeed, we have found that SOBAs are quite likely to disperse in a Hubble time.
However, MYCs with compact cores are expected to survive for a Hubble time (even
though their halos may disperse) and it should be to those to which the term 
should be applied exclusively. We also recommend that a neutral term like MYC be
applied to more distant unresolved objects with $\MVmax/ < -11$ or 
$\gtrsim\sci{3}{4}$ \Ms/ until their structure can be analyzed.

\subsection{The masses of NGC 1569-A and NGC 1705-I-A}

	\citet{HoFili96} measured the velocity dispersions of NGC 1569-A and 
NGC 1705-I-A by cross-correlating their $5000-6280$ \AA\ spectra with a K5-M0
supergiant template. They obtained values of $15.7 \pm 1.5$ km s\m/ and 
$11.4\pm 1.5$ km s\m/, respectively, and then used the half-light radii measured
by \citet{Meuretal95} to obtain masses of \scipm{3.3}{0.5}{5} \Ms/ and 
\scipm{8.2}{2.1}{4} \Ms/, respectively. However, we point out that the FOC 
and PC images used by \citet{Meuretal95} to measure \rh/ were obtained previous 
to the first HST service mission and that in their data the core of 
NGC 1705-I-A was strongly saturated. NGC 1705-I-A may well be termed the 
``great saturator'', since most of the prop. ID 7506 WFPC2 images were also 
affected by this problem.  However, as explained before, the saturation here was
only minor and non-saturated data was available to correct for it. Furthermore,
saturation affected only the central 9 pixels, where only $\approx 35\%$ of the 
total corrected flux is contained, so the measurement of \rh/ after the extra
flux addition in the central area should yield a correct value.

	Our value for \rh/ for NGC 1569-A (2.1 pc) is in a reasonable
agreement with those of \citet{Meuretal95} ($1.7\pm 0.2$ pc) and
\citet{OConetal94} (1.9 pc)\footnote{We are converting all values to an assumed
distance of 2.2 Mpc for NGC 1569 and 5.0 Mpc for NGC 1705.}. On the other hand,
our value for \rh/ for NGC 1705-I-A (5.3 pc) is higher than the ones measured
by those same authors ($0.9\pm 0.2$ pc and 3.4 pc, respectively), with the
difference being especially significant in the first case. Given the pre-COSTAR
character of their data and the strong saturation problems of the FOC images, 
we think that our values should be preferred.

	The largest contribution to the uncertainty in the measured masses is
the distance, which enters the calculation through \rh/. Using the values 
for the uncertainties in the distance given by \citet{DeMaetal97} for NGC 1569 
($\pm 0.6$ Mpc) and by \citet{OConetal94} for NGC 1705 ($\pm 2$ Mpc), we arrive 
at values for the masses of NGC 1569-A and NGC 1705-I-A of \scipm{3.6}{1.0}{5} 
\Ms/ and \scipm{4.8}{1.9}{5} \Ms/. These values clearly establish NGC 1569-A 
and NGC 1705-I-A as bona fide SSCs (i.e. globular cluster progenitors) and also
solve an apparent contradiction in the \citet{HoFili96} results: NGC 1569-A
appeared to be $\approx 4$ times more massive than NGC 1705-I-A while at the
same time having a similar $M_V$ (and, since NGC 1705-I-A appears to be older 
than NGC 1569-A, the latter one is actually dimmer in \MVmax/). Now the results
are consistent with both clusters having a similar mass-to-light ratio when
reduced to the same age. However, one last word of caution should be spoken:
since NGC 1569-A is now known to have a double core and the value derived here 
assumes spherical symmetry \citep{HoFili96}, its mass may be
slightly lower than what is published here, since some of the width of the line
may be caused by the orbital motion of one core around the other and not by 
random stellar motions.

\subsection{Early cluster evolution}

	We mentioned earlier that during most of the life of a cluster the two 
most important processes which determine its evolution are two-body evaporation 
and tidal disruption. In the first few million years, however, mass loss from
stellar winds and SNe and binary heating are expected to play a dominant role in
the case of SSCs. Even though the mass lost is a small fraction of the total 
mass ($\approx 4\%$), this mass comes from deep inside the potential well and 
produces a significant expansion of the cluster \citep{PorZetal99}. The core is
expected to expand by a factor of $\approx 2$ in the first 7 Myr, with the exact
value depending on the model and on the definition of radius chosen.

	We decided to test this prediction with our data by checking whether
there is a dependence of the average cluster radius with age. We included in our
sample all the SSCs with a weak halo and those SSCs with a strong halo which
showed a single non-elongated core (30 Doradus, NGC 2403-II, and NGC 4214-I-A).
In order to establish a meaningful comparison, we used \rq/ for the weak-halo
SSCs and we measured \re/ for the strong-halo SSCs. Since all strong-halo SSCs
have values of halo/(halo+core) luminosities $\gtrsim 0.5$, \re/ should be a
good approximation (or at least a good upper bound) for the value of \rq/ in
the absence of a halo. We
discarded the strong-halo SSCs with a double or elongated core because we are
interested in the evolution of the type of simple systems described by 
\citet{PorZetal99} with no external influences. The measured values of \re/ for 
30 Doradus, NGC 2403-II, and NGC 4214-I-A are 0.89 pc, 1.02 pc, and 1.05 pc,
respectively. The results are plotted as a function of age in
Fig.~\ref{sizeagecores}. The age errors are obtained from Table~\ref{results}
while the errors in radius are estimated by assuming a measurement error of 1/4
of a pixel and a distance error of 20\% (10\% in the case of 30 Dor).

	An apparent evolution is seen in Fig.~\ref{sizeagecores}. The average
radius for the five clusters with most-likely age less than 7 Myr is 0.99 pc 
while that for the five clusters with most-likely age greater than 7 Myr is 
1.71 pc. The effect may be somewhat stronger than what appears in 
Fig.~\ref{sizeagecores} if we consider that two of the points in the lower left
part of the diagram could actually be below their represented positions: NGC
4214-I-A may be more compact if it is located closer than 4.1 Mpc (see previous 
note in section 2.3); also, R136 contains only $\approx 10\%$ of the $U$ 
flux of 30 Doradus, so \re/ is clearly only an upper bound of what \rq/ would be
in the absence of a halo. We can then conclude that the predicted expansion is 
apparently observed, though more clusters need to be measured and better data
have to be obtained in order to confirm it. It is expected that a very young 
core like R136 will expand in the next few Myr until it reaches a size more 
similar to that of NGC 1705-I-A, which is also a typical size for a mature GC.

\section{SUMMARY}

	We have analyzed a sample of 27 nearby MYCs and we have confirmed the
dichotomy between SSC cores and SSC halos/SOBAs. SSC cores are compact objects
which have a good chance of lasting for a Hubble time, maybe retaining a part of
their halos but losing most of them due to tidal interactions with their host
galaxies. SOBAs are extended clusters which are expected to disperse rather
quickly, thus producing a significant contribution to the field population of 
their host galaxies. This dichotomy places restrictions on the role of 
coalescence as the main mechanism for producing very massive stars and leads to 
some suggestions regarding the classification of massive clusters. We have found
an interesting similarity between the morphologies of very young SOBAs and SSCs 
with strong halos on the one hand and that of the densest parts of galactic 
molecular clouds, suggesting that the first retain a memory of their previous
stage. Our data have
also enabled us to obtain new values for the masses of NGC 1569-A and NGC
1705-I-A and to verify the prediction that SSC cores should experience an
expansion during their first few Myr of existence due to mass loss from stellar
winds and SNe.

\acknowledgments

The author would like to thank Jennifer Wiseman, Georges Meylan, and an
anonymous referee for helpful comments and/or discussion. Support for this work 
was provided by NASA through grant GO-8163.01-97A from the Space Telescope 
Science Institute, Inc., under NASA contract NAS5-26555.

\bibliographystyle{apj}
\bibliography{general}

\begin{deluxetable}{llclll}
\tablecaption{Archival HST/WFPC2 data used for this work. The main filter is
the one used to measure the radial profile. The last column indicates whether
the cluster was observed using the PC or one of the WF chips.\label{data}}
\tabletypesize{\small}
\tablewidth{0pt}
\tablehead{\colhead{Galaxy} & \colhead{Clusters} & \colhead{Main filter} & 
\colhead{$V$ filter} & \colhead{Proposal ID(s)} & \colhead{PC/WF}}
\startdata
LMC & 30 Dor            & F336W & F555W & 5589,5114,8163 & PC+WF \\
M33      & NGC 595           & F336W & F547M & 5384      & PC \\
         & NGC 604           & F336W & F555W & 5237      & WF \\
I Zw 18  & I,II              & F336W & F555W & 5309      & PC \\
NGC 1569 & A,B,C             & F336W & F555W & 6423      & PC \\
NGC 1705 & I-A,I-B           & F380W & ---   & 7506      & PC \\
NGC 2403 & I-A,I-B,I-C,II    & F439W & F547M & 5383      & PC \\
         & IV                & F439W & F547M & 5383      & WF \\
NGC 4214 & I-A,I-B           & F336W & F555W & 6716      & PC \\
         & I-D,II-C,V,VI,VII & F336W & F555W & 6569      & WF \\
NGC 4449 & N-1,N-2           & F336W & F547M & 6716      & PC \\
NGC 5253 & I,IV,VI           & F336W & F547M & 6716,6524 & PC \\
\enddata
\end{deluxetable}

\begin{deluxetable}{lcrlrl}
\tablecaption{Input data. Extinctions and ages are calculated from a combination of literature data and integrated colors. The fourth column indicates the method used to determine the age.\label{input}}
\tabletypesize{\small}
\tablewidth{0pt}
\tablehead{\colhead{Cluster} & \colhead{$E(B-V)$} & \colhead{Age} & \colhead{Method} & \colhead{$d$} & \colhead{References} \\ 
 & & \colhead{Myr} & & \colhead{Mpc} & }
\startdata
30 Dor        & 0.35 & $ 2.0\pm 1.0$ &     CMD,UVS,OSF &  0.05 & 1,2,3       \\ 
NGC 595       & 0.33 & $ 4.0\pm 1.0$ &     CMD,UVS,OSF &  0.84 & 4,5         \\ 
NGC 604       & 0.20 & $ 3.5\pm 0.5$ & CMD,UVS,OSF,Neb &  0.84 & 6,7         \\ 
I Zw 18-I     & 0.05 & $ 3.5\pm 0.5$ &     CMD,OSF,Neb & 10.00 & 8,9         \\ 
I Zw 18-II    & 0.20 & $ 3.0\pm 2.0$ &          Neb,IC & 10.00 & 8           \\ 
NGC 1569-A    & 0.55 & $ 6.0\pm 4.0$ &           OS,IC &  2.20 & 10,11       \\ 
NGC 1569-B    & 0.55 & $11.0\pm 3.0$ &           OS,IC &  2.20 & 10,11       \\ 
NGC 1569-C    & 1.00 & $ 3.0\pm 2.0$ &      OSF,Neb,IC &  2.20 & 12,13       \\ 
NGC 1705-I-A  & 0.06 & $15.0\pm 5.0$ &          OSF,IC &  5.00 & 14,15       \\ 
NGC 1705-I-B  & 0.06 & $10.0\pm 8.0$ &              IC &  5.00 & 14,15       \\ 
NGC 2403-I-A  & 0.28 & $ 6.0\pm 4.0$ &          OSF,IC &  3.20 & 16          \\ 
NGC 2403-I-B  & 0.28 & $ 6.0\pm 4.0$ &          OSF,IC &  3.20 & 16          \\ 
NGC 2403-I-C  & 0.28 & $ 6.0\pm 4.0$ &          OSF,IC &  3.20 & 16          \\ 
NGC 2403-II   & 0.28 & $ 4.5\pm 2.5$ &          OSF,IC &  3.20 & 16          \\ 
NGC 2403-IV   & 0.28 & $ 4.5\pm 2.5$ &          OSF,IC &  3.20 & 16          \\ 
NGC 4214-I-A  & 0.03 & $ 3.5\pm 0.5$ &  UVS,OSF,Neb,IC &  4.10 & 6,17,18     \\ 
NGC 4214-I-B  & 0.35 & $ 3.5\pm 0.5$ &      OSF,Neb,IC &  4.10 & 6,17        \\ 
NGC 4214-I-D  & 0.02 & $ 9.0\pm 3.0$ &      OSF,Neb,IC &  4.10 & 6,17        \\ 
NGC 4214-II-C & 0.20 & $ 2.0\pm 1.0$ &         OSF,Neb &  4.10 & 6,17        \\ 
NGC 4214-V    & 0.03 & $11.0\pm 5.0$ &              IC &  4.10 & 17          \\ 
NGC 4214-VI   & 0.05 & $11.0\pm 5.0$ &              IC &  4.10 & 17          \\ 
NGC 4214-VII  & 0.03 & $11.0\pm 5.0$ &          OSF,IC &  4.10 & 17          \\ 
NGC 4449-N-1  & 0.25 & $11.0\pm 5.0$ &              IC &  3.90 & 19          \\ 
NGC 4449-N-2  & 0.25 & $ 3.0\pm 2.0$ &              IC &  3.90 & 19          \\ 
NGC 5253-I    & 0.05 & $11.5\pm 2.5$ &              IC &  4.10 & 20          \\ 
NGC 5253-IV   & 0.05 & $ 3.5\pm 0.5$ &      OSF,Neb,IC &  4.10 & 20          \\ 
NGC 5253-VI   & 0.05 & $11.0\pm 3.0$ &              IC &  4.10 & 20          \\ 
\enddata
\par {\scriptsize \begin{tabular}{lp{12.5cm}}
Age methods & CMD: Color-magnitude diagram, UVS: Ultraviolet spectroscopy, OSF: Optical spectral features (WR, Ca triplet), Neb: Nebular equivalent widths and/or structure, IC:  Integrated colors. \\
 References &  1: \citet{WalbBlad97};   2: \citet{MassHunt98};
  3: \citet{Barbetal01};   4: \citet{Malaetal96};
  5: \citet{MasHKunt99};   6: \citet{Maiz00};
  7: \citet{GonDPere00};   8: \citet{HuntThro95};
  9: \citet{Legretal97};  10: \citet{DeMaetal97};
 11: \citet{GonDetal97};  12: \citet{Huntetal00};
 13: \citet{Bucketal00};  14: \citet{OConetal94};
 15: \citet{HoFili96};    16: \citet{Drisetal99b};
 17: \citet{MacKetal00};  18: \citet{Leitetal96};
 19: \citet{Gelaetal01};  20: \citet{Calzetal97}.
  \\ 
\end{tabular}}
\end{deluxetable}

\begin{deluxetable}{lrrrrcrl}
\tablecaption{Results. \rq/, \rh/, and \rqqq/ are obtained from the measured angular sizes and the distances. Age-corrected absolute $V$ magnitudes are calculated from the measured apparent $V$ magnitudes and the data in Table \ref{input}. Two values of $M_{V,{\rm max}}$ are given, one for the whole cluster and another one for the area within 3 pc of the center.\label{results}}
\tabletypesize{\small}
\tablewidth{0pt}
\tablehead{\colhead{Cluster} & \colhead{$r_{1/4}$} & \colhead{$r_{1/2}$} & \colhead{$r_{3/4}$} & \colhead{$m_V$} & \colhead{$M_{V,{\rm max}}$} &  \colhead{$M_{V,{\rm max}}$} & \colhead{Notes} \\ 
 & \colhead{pc} & \colhead{pc} & \colhead{pc} & & \colhead{tot} & \colhead{$<3$ pc} & }
\startdata
30 Dor        &  3.3 &  9.2 &  15.5 &  8.0 & $-12.4$ & $-10.7\;\;\;$ & C,St         \\ 
NGC 595       & 14.4 & 26.9 &  38.0 & 14.3 & $-11.4$ & $ -8.5\;\;\;$ & SOBA         \\ 
NGC 604       & 18.4 & 28.4 &  44.3 & 12.7 & $-12.6$ & $ -8.3\;\;\;$ & SOBA         \\ 
I Zw 18-I     & 29.1 & 46.3 &  75.0 & 17.0 & $-13.2$ & $ -6.3\;\;\;$ & SOBA         \\ 
I Zw 18-II    & 27.8 & 50.4 &  68.9 & 18.4 & $-12.3$ & $ -6.8\;\;\;$ & SOBA         \\ 
NGC 1569-A    &  0.9 &  2.1 &   5.3 & 14.6 & $-14.1$ & $-13.5\;\;\;$ & C,Wk,Db      \\ 
NGC 1569-B    &  1.4 &  3.7 &   9.5 & 15.4 & $-14.1$ & $-13.2\;\;\;$ & C,Wk         \\ 
NGC 1569-C    &  1.0 &  2.9 &   4.2 & 17.1 & $-12.8$ & $-11.8\;\;\;$ & C,Wk,Db      \\ 
NGC 1705-I-A  &  1.5 &  5.3 &  20.5 & 14.7 & $-15.4$ & $-14.5\;\;\;$ & C,Wk,Nb      \\ 
NGC 1705-I-B  &  1.9 &  2.7 &   4.9 & 18.1 & $-11.6$ & $-11.2\;\;\;$ & C,Wk,Nb      \\ 
NGC 2403-I-A  & 11.0 & 20.6 &  31.1 & 16.0 & $-12.6$ & $ -9.6\;\;\;$ & SOBA         \\ 
NGC 2403-I-B  & 14.8 & 26.3 &  33.1 & 16.7 & $-12.0$ & $ -8.6\;\;\;$ & SOBA         \\ 
NGC 2403-I-C  &  9.9 & 19.6 &  33.4 & 17.6 & $-11.0$ & $ -7.4\;\;\;$ & SOBA         \\ 
NGC 2403-II   &  2.0 & 11.8 &  27.2 & 15.1 & $-13.4$ & $-11.9\;\;\;$ & C,St         \\ 
NGC 2403-IV   &  9.3 & 30.0 &  46.7 & 15.7 & $-12.7$ & $-10.4\;\;\;$ & C,St,Db      \\ 
NGC 4214-I-A  &  2.1 & 16.5 &  38.9 & 14.5 & $-13.7$ & $-11.9\;\;\;$ & C,St         \\ 
NGC 4214-I-B  & 21.5 & 33.0 &  51.8 & 15.6 & $-13.6$ & $ -9.5\;\;\;$ & SOBA         \\ 
NGC 4214-I-D  &  9.9 & 15.3 &  27.6 & 17.2 & $-12.3$ & $ -9.2\;\;\;$ & SOBA         \\ 
NGC 4214-II-C & 16.6 & 21.7 &  29.2 & 18.1 & $-11.4$ & $ -8.4\;\;\;$ & SOBA         \\ 
NGC 4214-V    & 52.5 & 83.9 & 127.1 & 16.0 & $-13.3$ & $ -8.2\;\;\;$ & SOBA         \\ 
NGC 4214-VI   & 20.9 & 35.9 &  58.5 & 18.1 & $-11.3$ & $ -7.4\;\;\;$ & SOBA         \\ 
NGC 4214-VII  & 24.8 & 40.4 &  56.9 & 17.0 & $-12.3$ & $ -7.7\;\;\;$ & SOBA         \\ 
NGC 4449-N-1  &  6.9 & 16.9 &  29.3 & 14.5 & $-15.4$ & $-13.0\;\;\;$ & C,St,Nb,El   \\ 
NGC 4449-N-2  &  2.2 &  5.8 &  10.6 & 16.4 & $-12.4$ & $-10.9\;\;\;$ & C,St,Nb,El   \\ 
NGC 5253-I    &  2.0 &  4.0 &   7.5 & 16.9 & $-12.4$ & $-11.9\;\;\;$ & C,Wk         \\ 
NGC 5253-IV   &  6.2 & 13.8 &  19.4 & 16.3 & $-11.9$ & $-10.4\;\;\;$ & C,St,Db      \\ 
NGC 5253-VI   &  1.7 &  3.1 &   6.1 & 18.0 & $-11.3$ & $-11.1\;\;\;$ & C,Wk         \\ 
\enddata
\par {\scriptsize \begin{tabular}{ll}
 Notes & C: Compact cluster, SOBA: Scaled OB Association. \\
 & Wk,St: Weak ($<$40\%), Strong ($>$40\%) halo/total ratio. \\
 & Db: Double core ($\lesssim$ 7 pc separation), Nb: Nearby ($\lesssim$ 30 pc) cluster present. \\
 & El: Elongated core. \\
\end{tabular}}
\end{deluxetable}


\addtocounter{figure}{-1}

\begin{figure}
\plotone{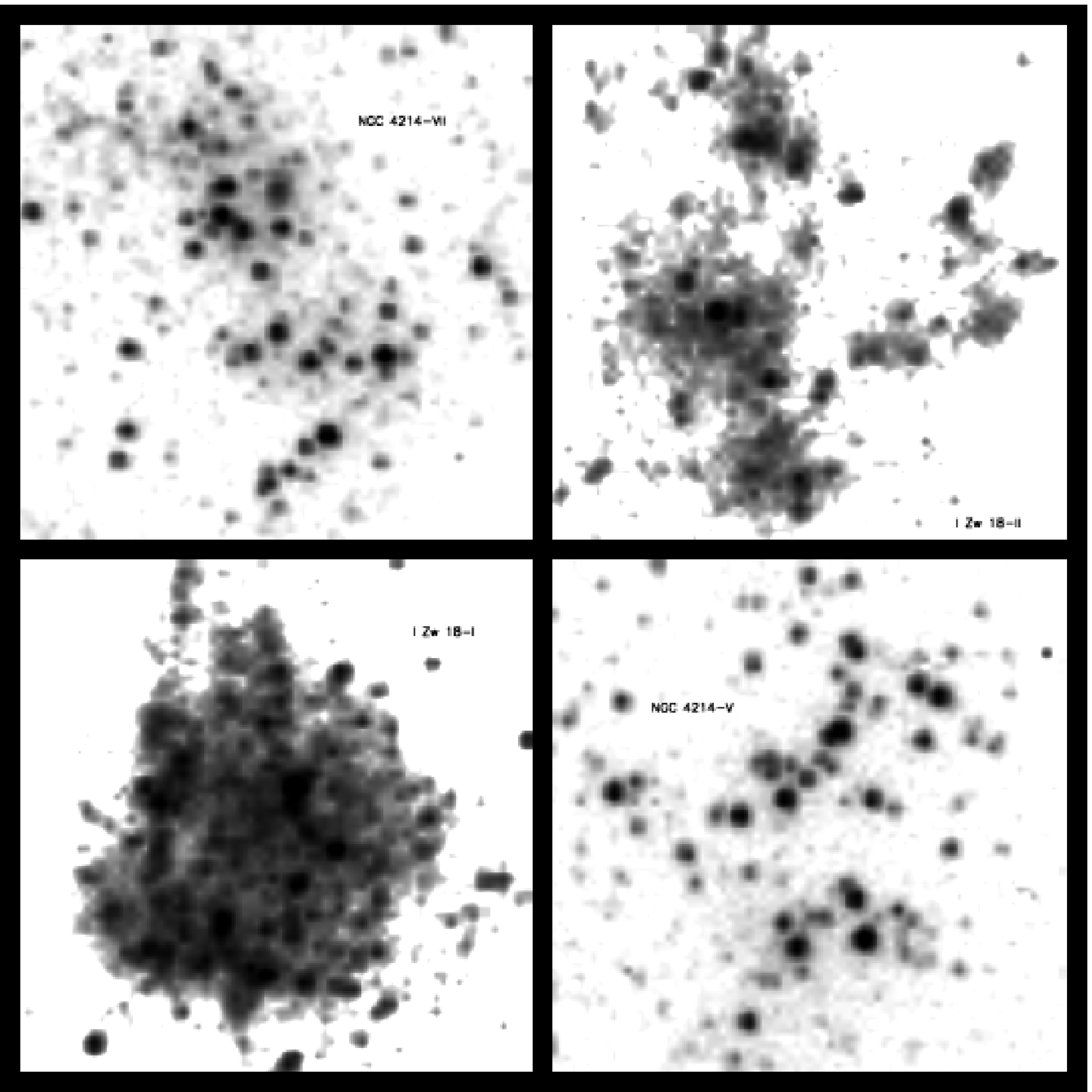}
\caption{(b) Continued. The field sizes are 150~pc~$\times$~150~pc.} 
\end{figure}

\begin{figure}
\plotone{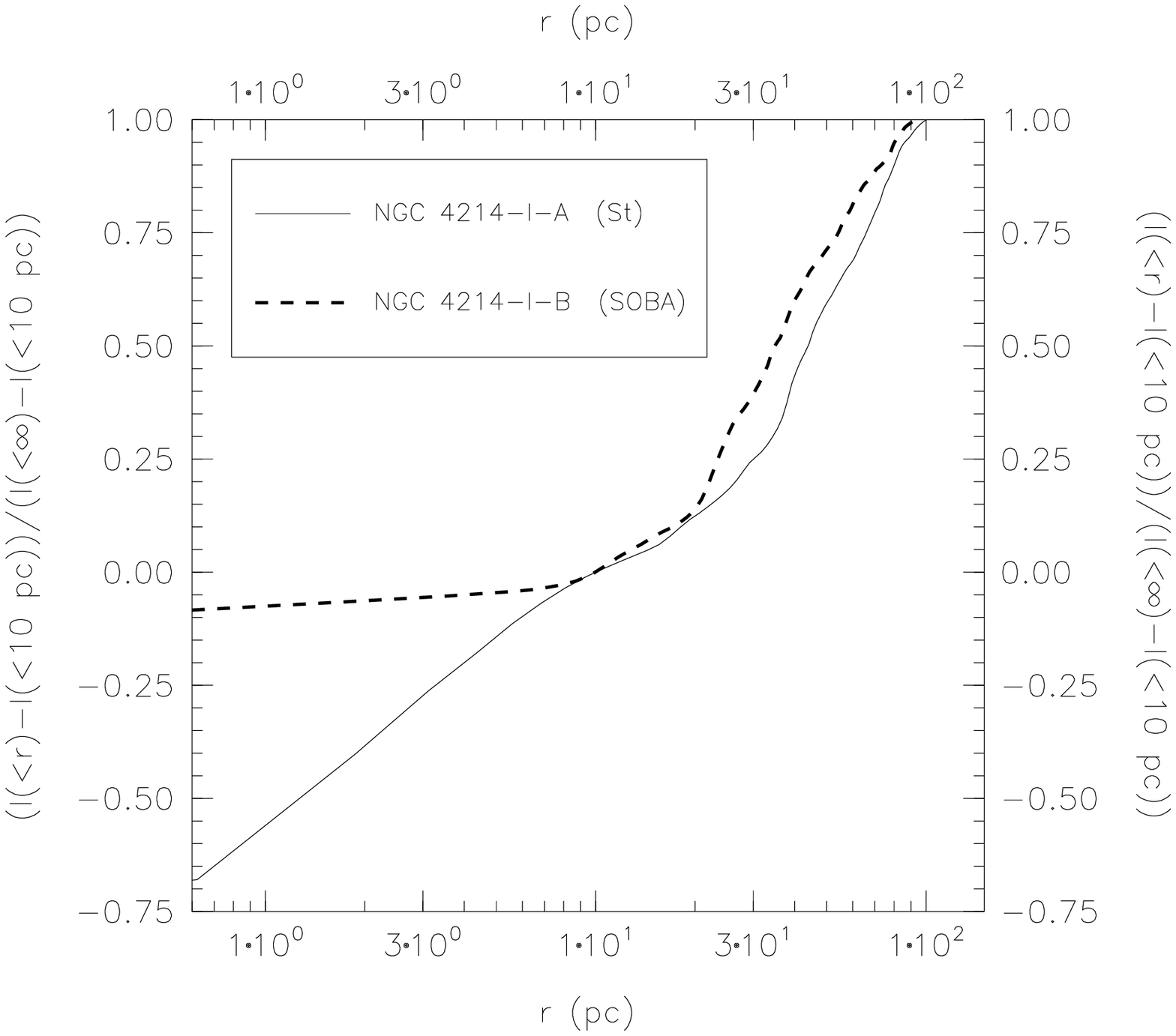}
\caption{Intensity profiles for a compact cluster with a strong halo and for a
SOBA. Note the similarity for $r > 10$ pc.}
\label{compintegrint}
\end{figure}

\begin{figure}
\plotone{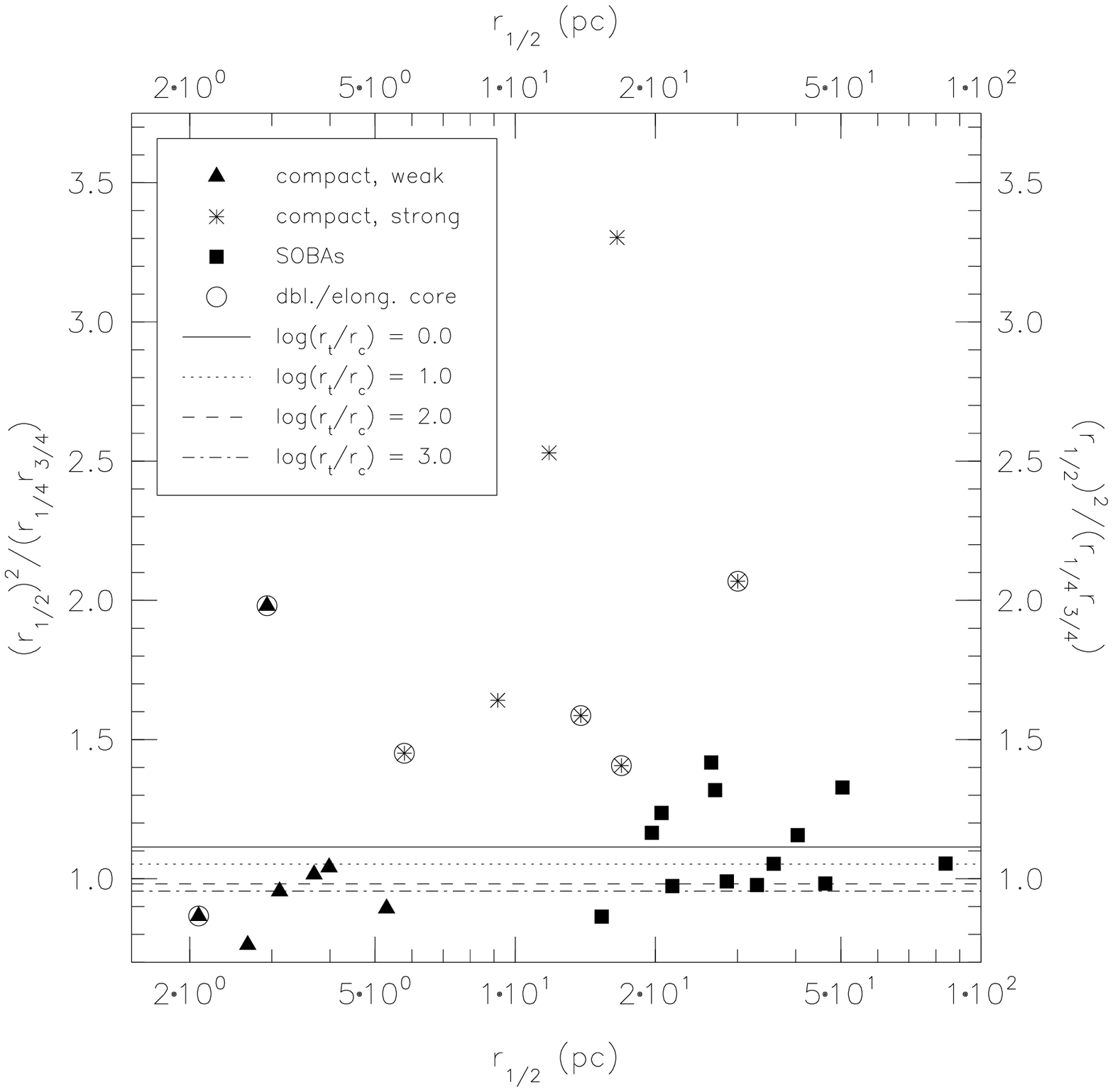}
\caption{$\alpha \equiv \rh/^2/(\rq/\rqqq/)$ as a function of \rh/ plot 
which shows the different regions occupied by compact clusters with a weak halo,
compact clusters with a strong halo, and SOBAs. Compact clusters with a double 
or elongated core are marked with a circle around their symbol. The horizontal
lines indicate the expected values for different King models.}
\label{classifplot}
\end{figure}
 
\begin{figure}
\plotone{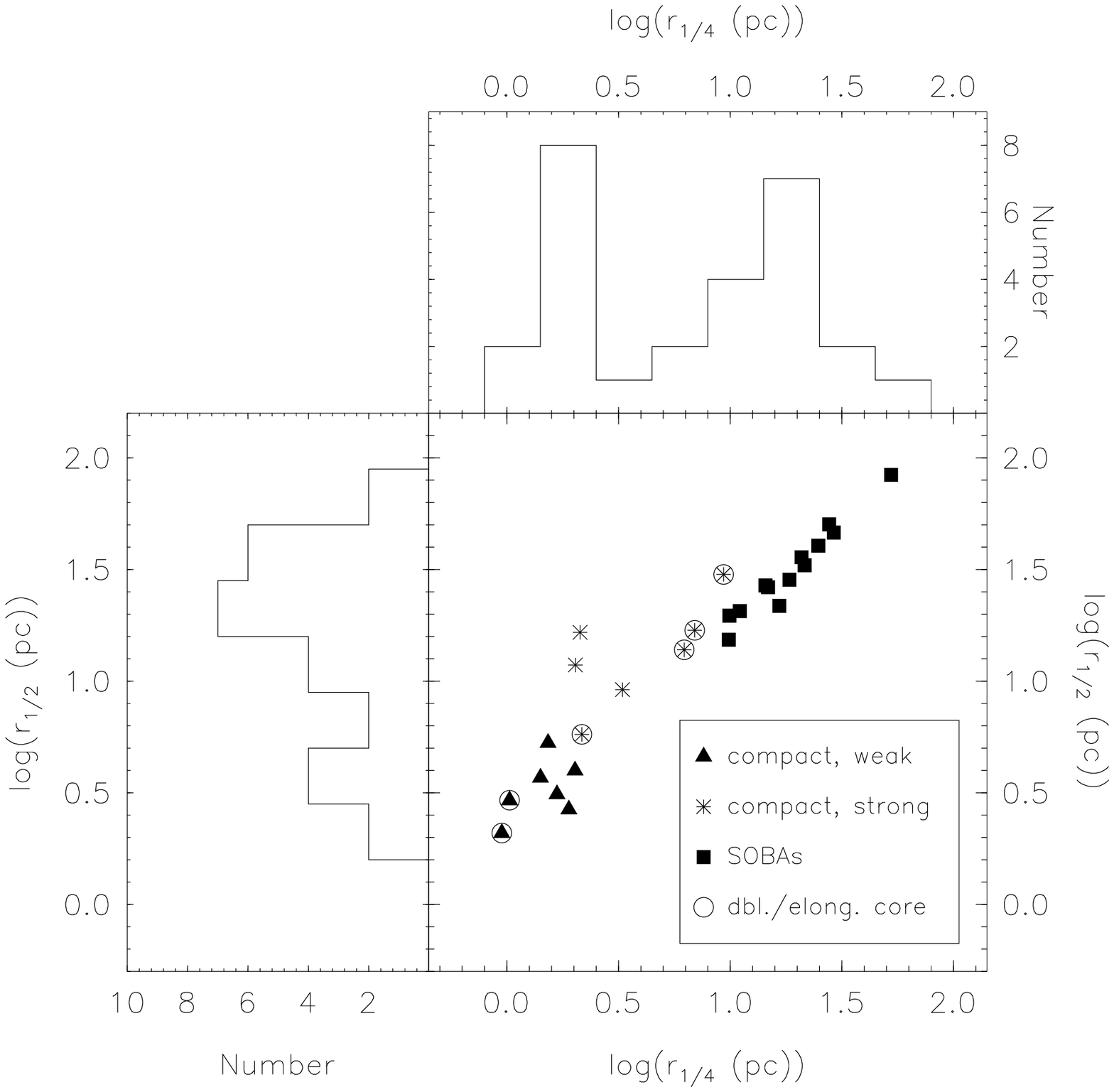}
\caption{Half-light radius vs. quarter-light radius plot for the 27 clusters in
the sample. Compact clusters with a double or elongated core are marked with a 
circle around their symbol. The top and left plots are the histograms for each 
axis.}
\label{r14r12}
\end{figure}
 
\begin{figure}
\plotone{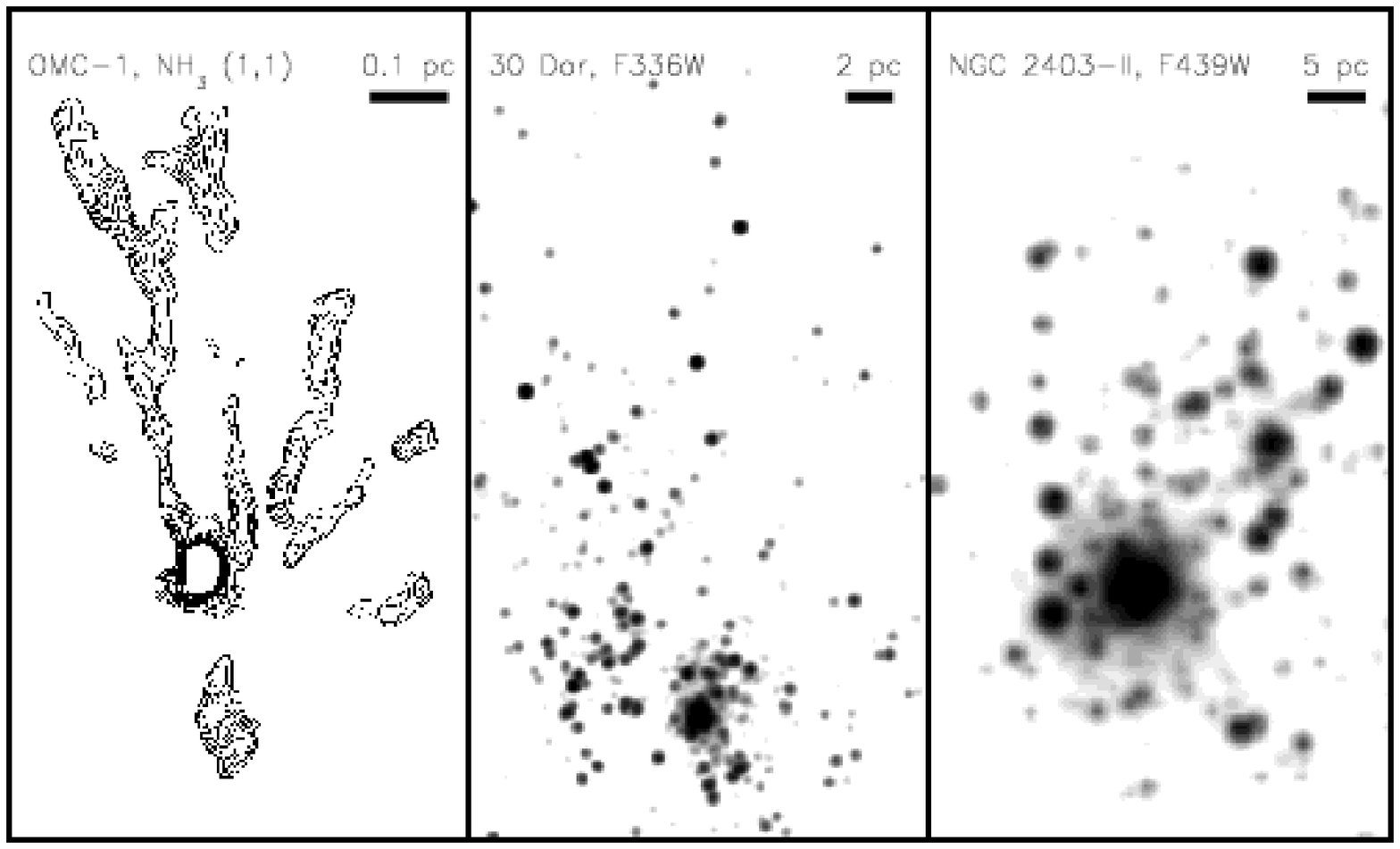}
\caption{A comparison between the structures observed in the dense material of a
galactic star-forming molecular cloud (OMC-1, adapted from \citealt{WiseHo98}) 
and in two of the compact clusters with strong halos in our sample. Even though
the scales are quite different, in both cases the same core + quasiradial 
filamentary structure is observed. Since structures in molecular clouds are
apparently hierarchical, it appears likely that the ``chains of stars'' in the
cluster halos are a consequence of the original mass distribution of the parent 
cloud.}
\label{starchains}
\end{figure}

\begin{figure}
\plotone{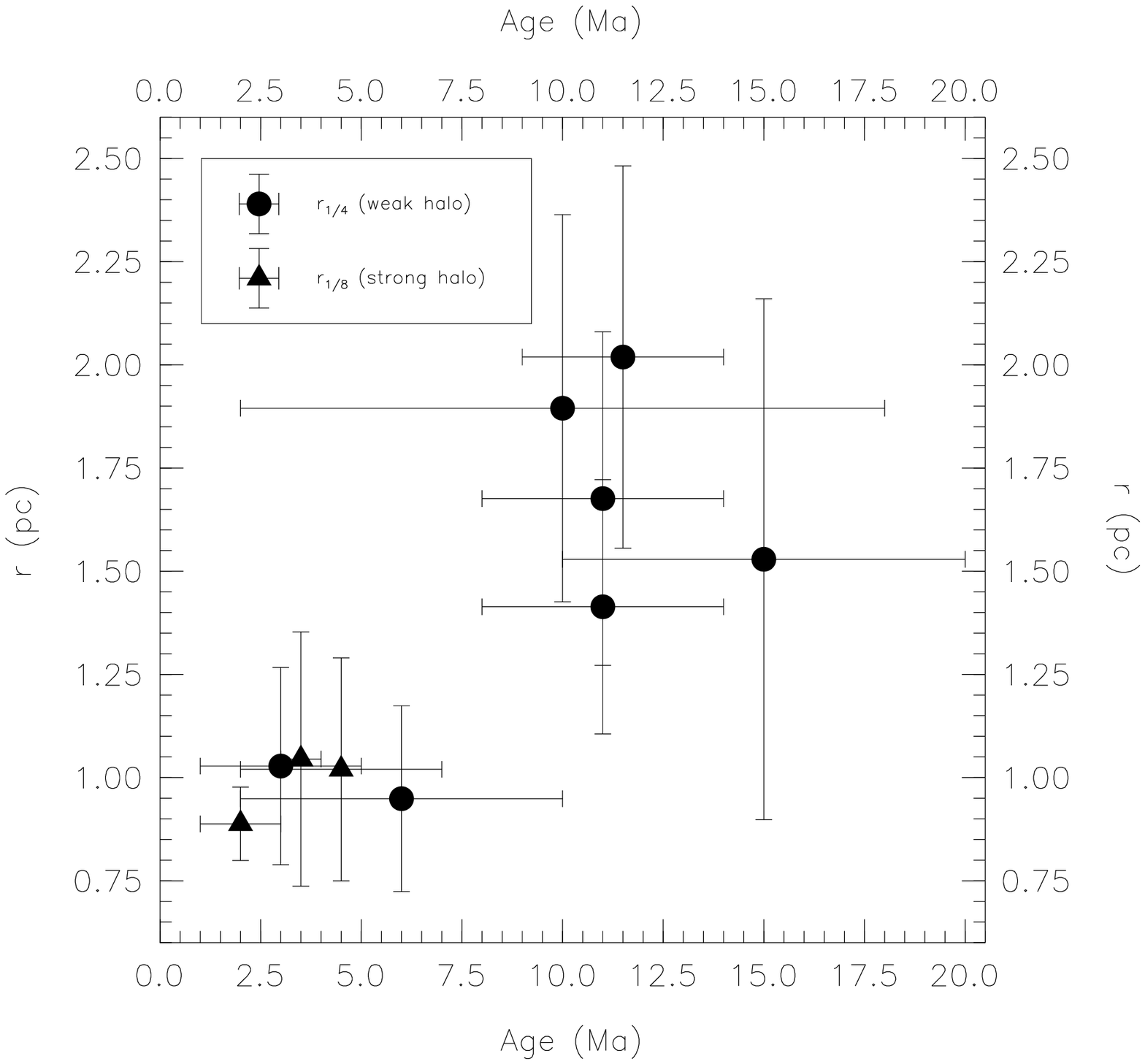}
\caption{Radius as a function of age for the compact clusters in the sample. 
For objects with weak halos, \rq/ is the plotted radius, while for objects with 
strong halos it is \re/. For clusters with strong halos only the cases in which
the core is not double or elongated are shown.}
\label{sizeagecores}
\end{figure}

\end{document}